\documentstyle{l-aa}
\begin{document}
\thesaurus{09.04.1;10.19.1;04.01.1;03.13.6;10.15.1}

\title{Comparison of two different extinction laws with 
Hipparcos observations \thanks{Based in part on observations made 
with the ESA Hipparcos astrometry satellite}} 

\author{B. Chen \inst{1} \and J.L. Vergely \inst{2} \and B. Valette 
\inst{3}  \and G. Carraro \inst{4,5}}                      

\offprints{bchen@mizar.am.ub.es}

\institute{Departament d'Astronomia i Meteorologia,
           Universitat de Barcelona,
           Avda. Diagonal 647,
           E08028, Barcelona, Spain
\and Observatoire de Strasbourg, 11, rue de l'Universit\'e, 
F-67000, Strasbourg, France
\and Institut de Physique du Globe de Paris, F-75007, Paris, France
\and  Department of Astronomy, Padova University, Vicolo dell'Osservatorio
5, I-35122 Padova, Italy
\and SISSA/ISAS, via Beirut 2, I-34013, Trieste, Italy
}
\date{Received January 5, 1998; Accepted April 28, 1998}

\maketitle

\begin{abstract}

Interstellar absorption in the galactic plane is highly variable 
from one direction to another.
In this paper colour excesses and distances from a new  open cluster sample 
are used to investigate the spatial distribution of the
interstellar extinction. An inverse method (Tarantola \& Valette, 1982) 
is used to construct the
extinction map in the galactic plane below $|b| < 10^{o}$.   
The $A_{v} (r,l) $ diagrams are  compared with those derived from 
individual stars (Arenou et al. 1992, Neckel \& Klare 1980).  
An analytic expression for the interstellar extinction as a function of 
galactic longitude and distance in the solar neighborhood  is given.
The comparison of the model predictions with Hipparcos observations
 in the 4-dimensional space of ($V$, $B-V$, $H_v$, $r$) 
shows that our extinction model provides a better fit to the data.  
However, a new and more detailed extinction model  
is still lacking. 


\end{abstract}

\section{Introduction}

The light extinction caused  by the interstellar medium in the solar
neighborhood alters the apparent brightness of the stars and thus
their intrinsic properties (absolute magnitude,
intrinsic colour, etc.). For this reason, it is crucial to know
the effects of the interstellar medium on the radiation we
receive from the stars, and to build models up which
reasonably predict the amount of extinction in a given direction.

Bahcall \& Soneira (1980)  have used Sandage's extinction law (1972)
in their model of the Galaxy, and compared model predictions  
with star count observations in 17 fields. They found that,  at  
$\mid$ b $\mid$ $>$ 10$^{o}$,  
the  Sandage absorption law is a good approximation to observations. 
However, at low galactic latitudes,  
the absorptions are patchy and poorly  known.  
In the context of our study on stellar kinematics from
Hipparcos observations and open cluster (Chen et al, 1998),
we need a reliable extinction model at low galactic latitudes.

Neckel \& Klare (1980)  derived extinctions and distances 
for more than 11000 stars and  investigated the spatial 
distribution of the interstellar extinction at $\mid$ b $\mid$  $<$ 7.6$^{o}$. 
Their results are shown in a series of diagrams. 
Using a larger sample (about 17000 stars) 
with MK spectral types and photoelectric photometry, Arenou et al. (1992)
(hereafter AGG92) constructed an extinction model in which  
the sky was divided into 199 cells. In each cell, a quadratic 
relation was adopted.
This extinction model will be the reference in this paper. 
Hakkila et al. (1997) have developed a computerized model of 
large-scale visual interstellar extinction. The code merges several 
published studies (FitzGerald 1968; Neckel \& Klare 1980; 
Berdnikov \& Pavlovskaya 1991; AGG92; etc.), and can be used for making 
corrections to individual observations and for correcting statistical sample.
With extinctions determined from Str\"{o}mgren photometry and parallaxes from 
Hipparcos for a sample of 3799 stars, Vergely et al (1997) have constructed 
a three dimension  distribution of absorbing matter in the solar neighbourhood. 
However, all these authors use similar dataset (stars along with their 
corresponding colors,  magnitudes, and spectral/morphological classification 
types) so that the results are not statistically independent (Hakkila et al. 1997).

In this paper, we are going to use open clusters rather than individual
stars to investigate the distribution of interstellar extinction, 
and compare our results with those  found in the literature. 
The reason for this investigation is twofold.
First, the advent of CCDs
with their sensitivity and linearity has allowed to obtain
deep photometry of open clusters (Carraro \& Ortolani, 1994, Kaluzny
1994). 
Second, distances and reddening
for open clusters are more reliable than those 
from individual stars, simply because one is dealing with a group of stars
and therefore the results are less 
sensitive to individual errors.
Thus the open clusters system provides us with
another way to obtain informations on A$_{v}$ as a function of the 
galactic longitude $l$ and distance $r$.

Pandey \& Mahra (1987) firstly  used the open cluster catalogue of 
Janes \& Adler (1982) to investigate the spatial distribution of 
interstellar extinction within 2 kpc. 
Their results are presented in graphical form. 
No analytical
expression for the
interstellar extinction has been presented.
In this paper, we have used a new open cluster database and  
an inverse method (Tarantola \& Valette, 1982) to   
derive an analytical expression of the
interstellar extinction in the solar vicinity, and compared 
different 
extinction models with 
Hipparcos observations. 
In section 2 we present the data used in this paper. 
In section 3 we briefly explain the AGG92's model and compare it with 
our open cluster data. In section 4, we construct a new extinction model 
from open cluster system. 
In section 5 the predictions from AGG92's extinction model 
and our extinction model are compared with the Hipparcos data. 
The main conclusions are summarized in section 6.

\section{The Data}

The main source of the data in this investigation is  
the Open  Cluster Data Base (Mermilliod 1992).
In this database,  
the reddening and distance for each cluster were determined from 
a system of weights corresponding to the precision of open cluster data. 
412 open clusters with reddening and distance are listed in the 
database. 
In addition to this data, 
from the list of open clusters, we have checked each cluster which 
has no information about reddening and distance from ADS database 
to see whether there are new observations published 
after 1992. Finally, we have found 22 such clusters, mostly
due to the observations from Padova and Boston group (Friel, 1995).
Thus, the present study is based on the observed distances and colour 
excesses of 434 open clusters. The ratio of total to selective 
absorption, $R$=3.1, was adopted to obtain total absorption from selective   
absorption $E(B-V)$.  
In Figure 1 we show the distribution of the open clusters  
in the ($l$, $b$) plane of galactic longitude and galactic latitude. 
We can see that most  open clusters (91\%) lie 
on the galactic plane ( $\mid$ b $\mid$ $\le$ 10$^{o}$ ), and 
that the distribution of open clusters 
is not homogeneous in galactic longitude ($l$). In the direction between  
$l$= 30$^{o}$ and $l$ = 90$^{o}$, and between $l$ = 145$^{o}$ and 
$l$ = 220$^{o}$, fewer open clusters are found.   

 \begin{figure}
      \mbox{}
      \vspace{10cm}
     \includegraphics{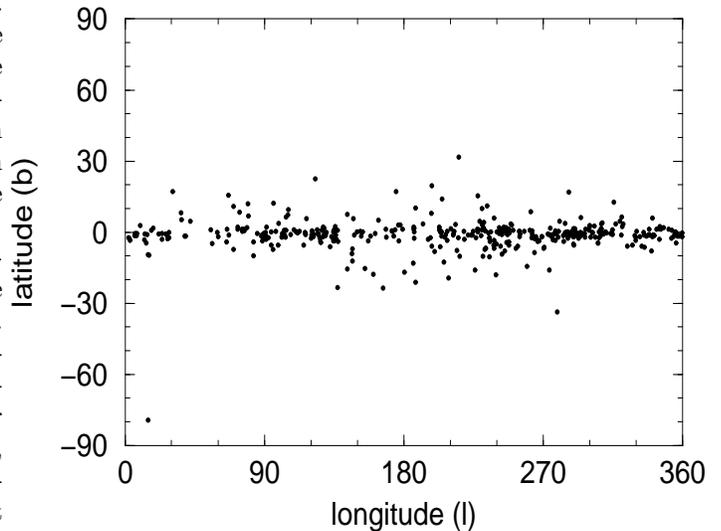}
  \caption[]{The distribution of the open clusters in the plane of
   galactic longitude and latitude}
   \end{figure}

In Figure 2 we show 
the histogram of the distance $r$. 
In the Open Cluster Data Base (Mermilliod, 1992), the errors of the 
extinction and distance are not given.  
We suppose that the error $\sigma_{Av}$
can be written :

\begin{equation}
\sigma_{Av(obs)}=\Delta Av(obs)_{cal} + \Delta Av(obs)_{r}
\end{equation}

Where $\Delta Av(obs)_{cal}$ is the
calibration error approximately equal to 0.1 mag
and $\Delta Av(obs)_{r}$ the extinction
error due to the distance error.
As a first approximation we consider that the distance determination
and $Av(obs)$ are uncorrelated, and that the relative distance error is 
0.25. Then, we have    
\begin{equation}
\sigma_{Av(obs)}=0.1+0.25 Av(obs)
\end{equation}

 \begin{figure}
      \mbox{}
      \vspace{9cm}
     \includegraphics{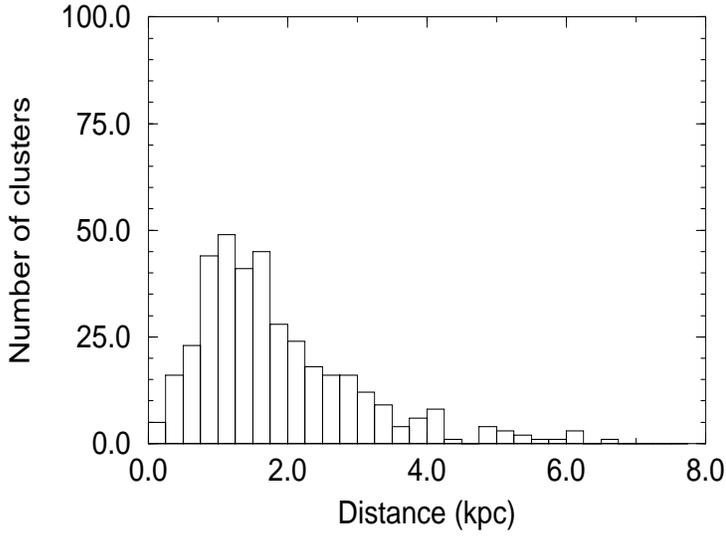}
  \caption[] { Histogram of distance $r$  to the Sun of the open
clusters}
   \end{figure}

 From Hipparcos 
parallaxes,  Mermilliod et al (1997) and van Leeuwen \& Hansen Ruiz (1997)
claimed that the distance of young Pleiades cluster is 10% 
$\sim$ 15\% smaller than its photometric distance. In order to 
check whether a systematic bias exists on all determinations of open 
clusters from photometric methods, we have compared the 
open cluster distance 
derived from Hipparcos data (Robichon et al. 1997; Mermilliod et al. 1997) 
with those in Mermilliod's catalogue. Figure 3 shows the distance derived 
from Hipparcos parallaxes ($r_{1}$) against those in Mermilliod's open 
cluster database ($r_{2}$), 
no systematic trend has been found.

 \begin{figure}
      \mbox{}
      \vspace{10cm}
     \includegraphics{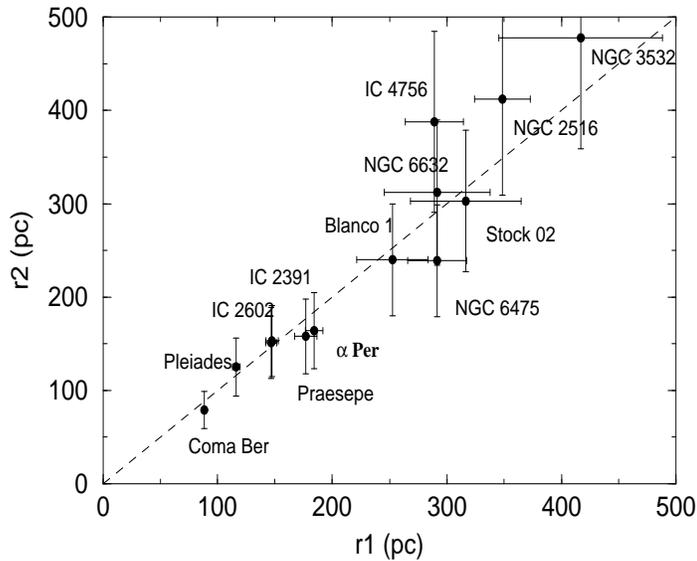}
  \caption[] { The open cluster distance derived from Hipparcos data ($r1$) 
against those in Mermilliod's open cluster database ($r2$)}
   \end{figure}

\section{AGG92's  extinction model}

AGG92 have used INCA database (G\'omez et al. 1989;
Turon et al. 1991) to
construct a tridimensional model of the galactic interstellar
extinction. For each star the distance $r$ to the sun and the visual
extinction $A_v$
have been computed from MK spectral type and photoelectric colour index.

 \begin{figure}
      \mbox{}
      \vspace{10cm}
     \includegraphics{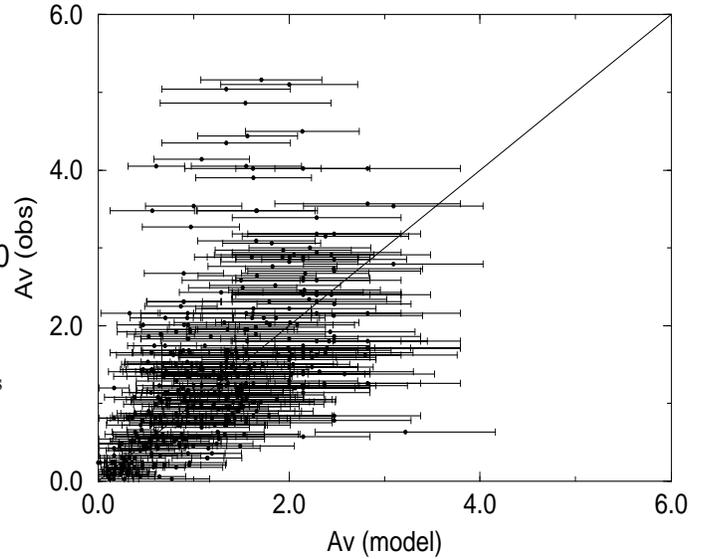}

  \caption[]{ The observed extinction A$_v$(obs) against that derived

from AGG92's model A$_v$ (model) for our open cluster sample}
   \end{figure}

 \begin{figure*}
      \mbox{}
      \vspace{17cm}
     \includegraphics{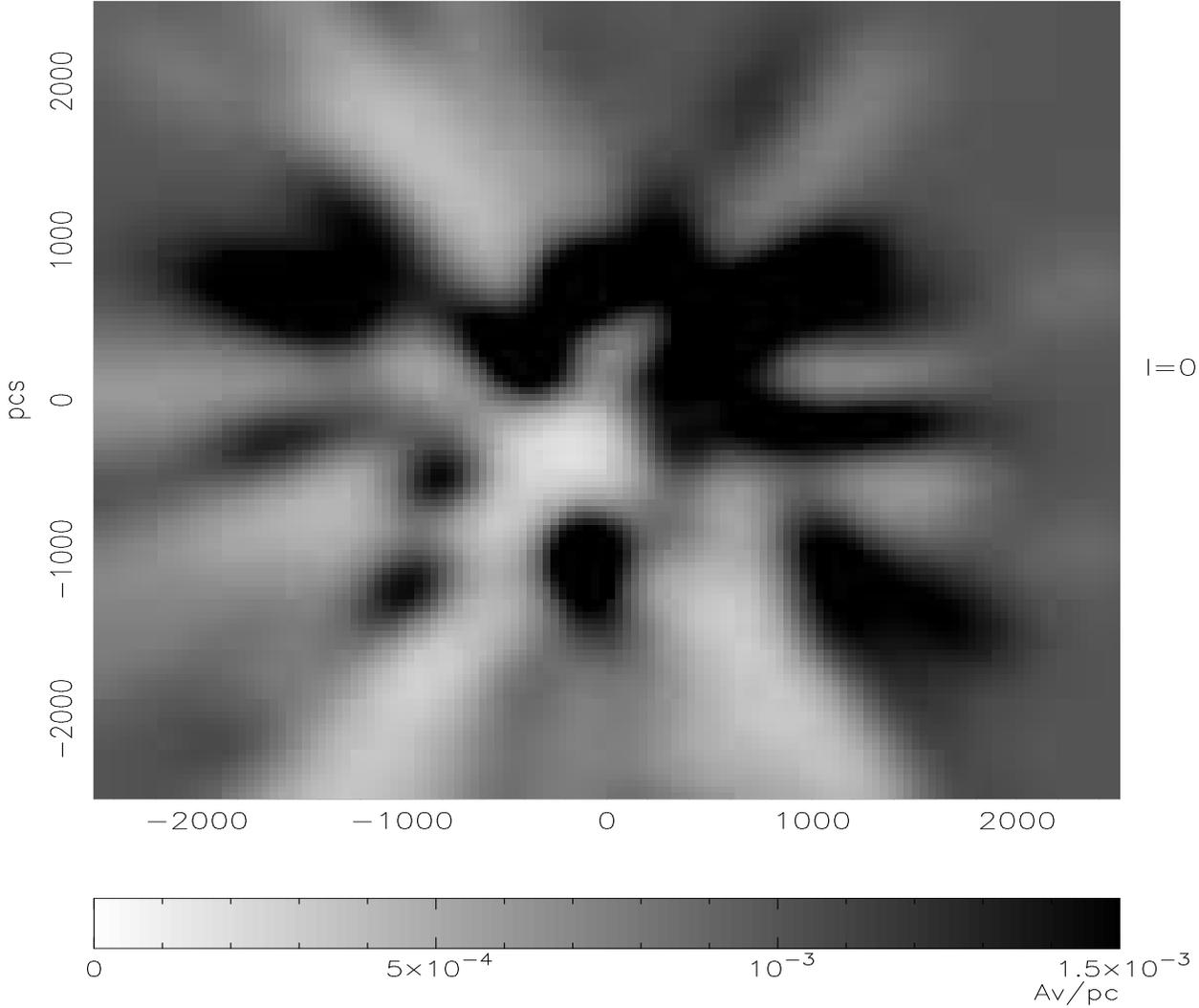}
  \caption[]{ Extinction structures in the Galactic plane}
   \end{figure*}

The sky was divided in 199  cells.
In each cell, based on the work of Neckel \& Klare (1980), a
quadratic relation was adopted:

\begin{equation}
A_{v} = \alpha r + \beta r{^2}
\end{equation}

 \begin{figure*}
      \mbox{}
      \vspace{8cm}
     \includegraphics{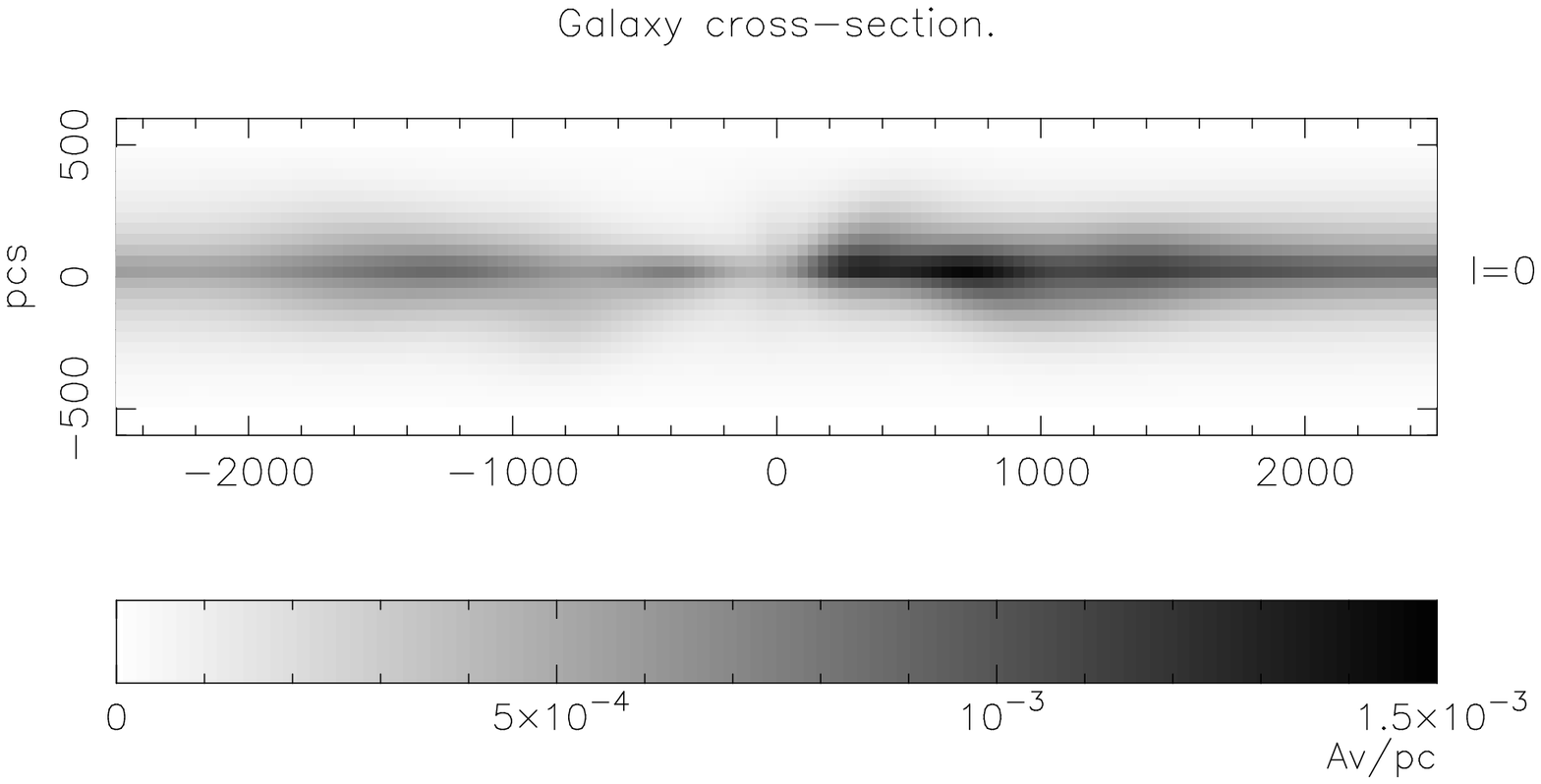}
  \caption[]{ Extinction structures in the cross section of the galactic plane} 
   \end{figure*}

 \begin{figure}
      \mbox{}
      \vspace{10cm}
     \includegraphics{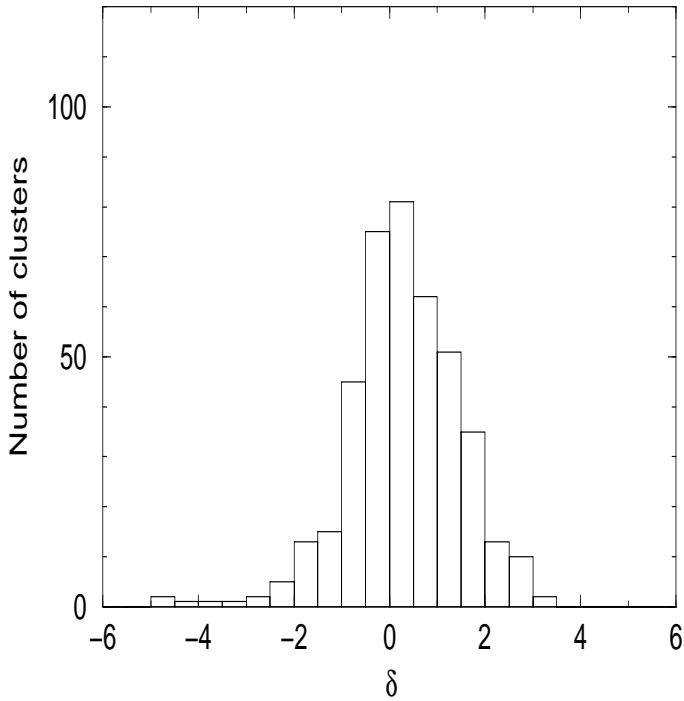}

  \caption[]{ Histogram of $\delta$ value
($\delta = \frac{Av(obs)-Av(model)}
{\sigma_{Av(obs)}}$).}
   \end{figure}

 \begin{figure}
      \mbox{}
      \vspace{9.5cm}
     \includegraphics{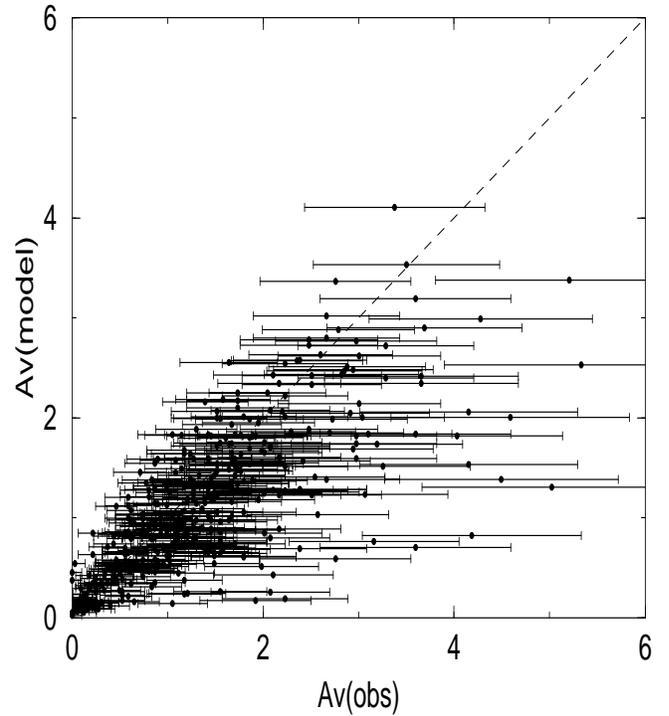}

  \caption[]{ The extinction derived from our model Av(model) against 
 the observed one Av(obs).}  
   \end{figure}

 \begin{figure*}

      \mbox{}
      \vspace{22cm}
     \includegraphics{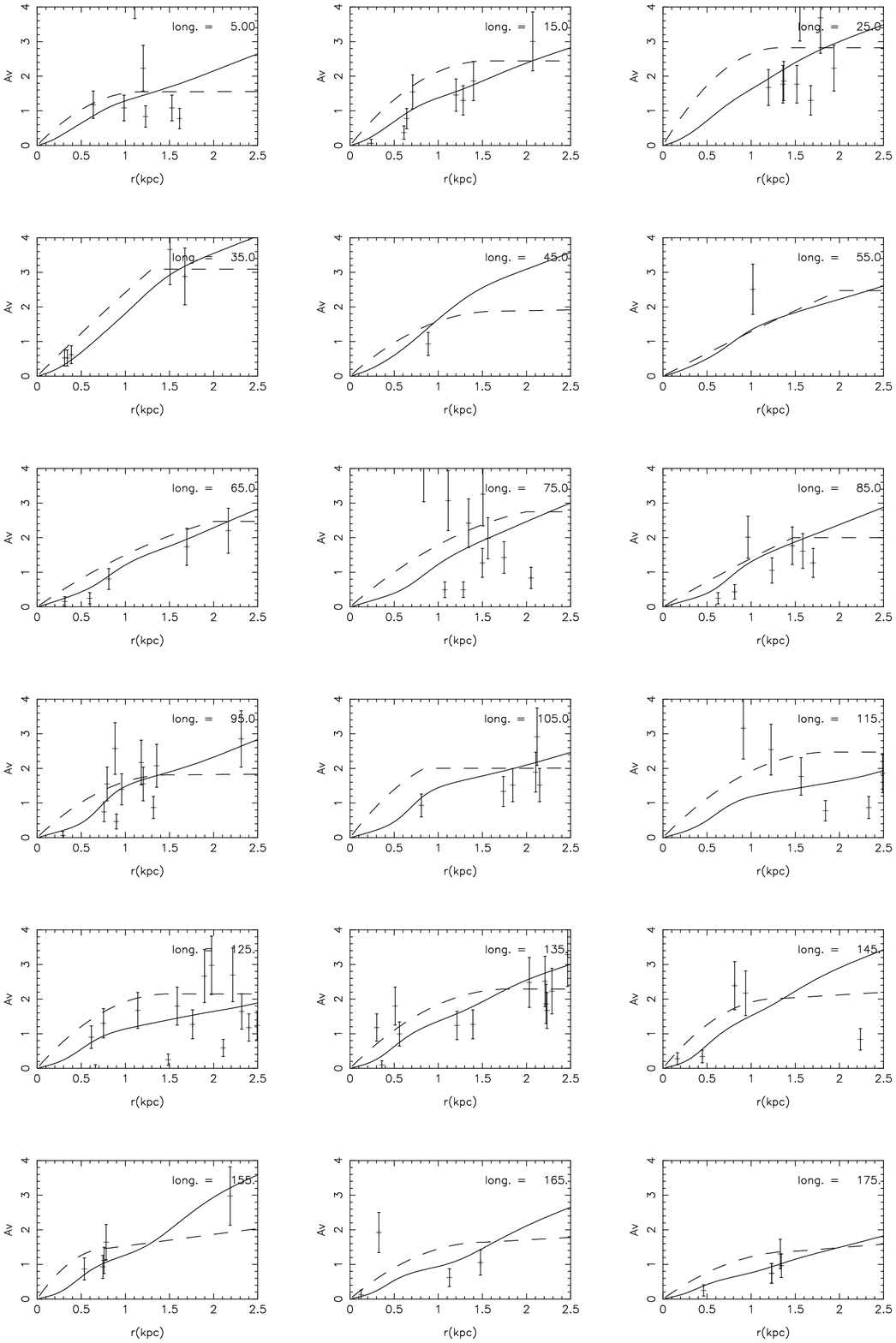}

  \caption[]{The extinction as a function of distance for each of the 36 cells.
    The solid lines are the results in this study from the inverse method, 
    the dashed lines are the results from  Arenou et al. (1992).
    The crosses (+) are the  
    open cluster sample.}
   \end{figure*}

\setcounter{figure}{8}
 \begin{figure*}

      \mbox{}
      \vspace{22cm}
     \includegraphics{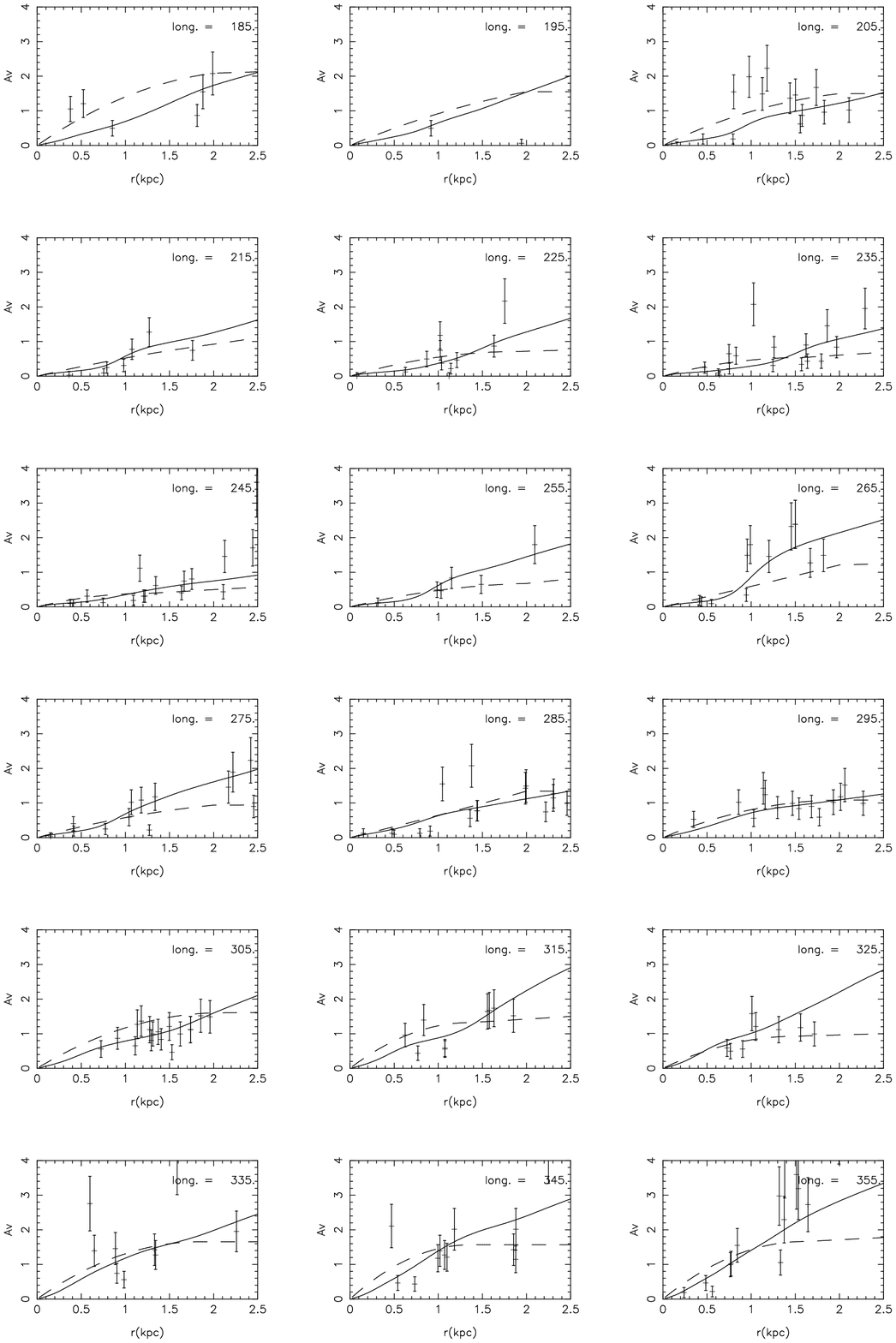}

  \caption[]{(continued)The extinction as a function of distance for 
   each of the 36 cells.
    The solid lines are the results in this study from the inverse method, 
    the dashed lines are the results from  Arenou et al. (1992).
    The crosses (+) are the 
    open cluster sample.}
   \end{figure*}

For each cell an analytic expression of the interstellar
monochromatic extinction at V-magnitude A$_v$($r$,$l$,$b$) was obtained.
We have used AGG92's model to  derive the extinction 
at the position of each open cluster in our sample. In Figure 4, 
we plot the observed extinction $A_v$(obs) against that 
derived from AGG92's model $A_v$(model). It can be seen that 
AGG92's model underestimates large extinction value. 
Since AGG92 used only stars brighter than V = 13 mag, 
as they stated, their model cannot be extrapolated farther than 1 kpc.
From Figure 4 we have also found that AGG92's 
results are higher than the observations 
for small distance ($r < 500$ pc). AGG92 pointed out that a slight 
overestimation of $A_v$ for small distance is expected in their model, 
this is consistent with what we have noticed from our open cluster sample.      
In the following 
sections, we use open clusters  
with $\mid$ b $\mid$ $\le$ 10$^{o}$ to derive the interstellar extinction 
at low galactic latitudes and construct a new extinction model 
in solar neighbourhood.

\section{Using open clusters to derive the interstellar extinction 
at low galactic latitudes}

In order to determine the opacity in each point of the space
in a circle of 2500 pc around the Sun, we utilized an inverse method
(Tarantola \& Valette, 1982; Valette \& Vergely, in preparation)
which for the first time has been applied to detect extinction
structures in the Hipparcos data (Vergely et al., 1997).\\

\begin{table*}
  \caption[]{\em The coefficients of the extinction curves in different longitudes}
%  \label{tab:table}
  \begin{center}
    \begin{tabular}{ccccc}
      \hline 
      Range of $l$  & $a_1$ & $a_2$  & $a_3$ & $a_4$ \\
      (degrees)   & (mag.kpc$^{-1}$) 
   & (mag.kpc$^{-2}$)   & (mag.kpc$^{-3}$)  & (mag.kpc$^{-4}$) \\
      \hline 
  0-10&     0.752&    0.610&    0.090&   -0.387\\
  10-20&    0.500&    3.012&   -3.659&    1.378\\
  20-30&    0.245&    5.468&   -7.721&    3.432\\
  30-40&    0.156&    5.996&   -9.214&    4.478\\
  40-50&    0.300&    3.753&   -5.793&    2.893\\
  50-60&    0.490&    9.364&   -1.023&    0.476\\
  60-70&    0.740&   -1.825&    3.780&   -1.622\\
  70-80&    0.943&   -3.715&    6.975&   -2.682\\
  80-90&    0.747&   -2.954&    6.672&   -3.190\\
  90-100&   0.598&   -2.604&    7.809&   -4.593\\
  100-110&  0.442&   -1.745&    7.066&   -4.558\\
  110-120&  0.235&    0.201&    2.681&   -2.111\\
  120-130&  0.065&    1.821&   -0.625&   -0.308\\
  130-140& -0.103&    3.096&   -2.055&    0.248\\
  140-150& -0.248&    4.273&   -3.300&    0.440\\
  150-160& -0.189&    4.592&   -4.951&    1.556\\
  160-170& -0.033&    4.066&   -5.594&    2.389\\
  170-180&  0.091&    3.101&   -4.644&    2.158\\
  180-190&  0.163&    2.070&   -3.064&    1.476\\
  190-200&  0.217&    1.084&   -1.410&    0.698\\
  200-210&  0.286&    0.189&   -0.117&    0.392\\
  210-220&  0.307&   -0.227&    0.434&    0.288\\
  220-230&  0.259&   -0.160&    0.649&   -0.266\\
  230-240&  0.201&    0.192&    0.123&   -0.128\\
  240-250&  0.198&    0.258&    0.160&   -0.136\\
  250-260&  0.274&   -0.082&    0.729&   -0.207\\
  260-270&  0.313&   -0.016&    0.187&    0.374\\
  270-280&  0.288&    0.386&   -0.462&    0.361\\
  280-290&  0.313&    0.535&   -0.650&    0.395\\
  290-300&  0.381&    0.398&    0.079&   -0.117\\
  300-310&  0.280&    1.661&   -1.839&    0.643\\
  310-320&  0.120&    3.513&   -5.033&    2.260\\
  320-330& -0.046&    5.282&   -7.844&    3.611\\
  330-340& -0.031&    5.477&   -7.961&    3.546\\
  340-350&  0.465&    2.076&   -2.138&    0.871\\
  350-360&  0.961&   -1.226&    3.345&   -1.648\\
      \hline \\
      \end{tabular}
  \end{center}
\end{table*}

In this study, we have assumed that the extinction follows a linear
law :
$$
E_i=\int_0^{R_i} \rho_{op}(l_i,b_i,r) dr
$$
where $\rho_{op}(r)$ is the opacity at the point $r$ in the
V-band (in mag/pc).
$E_i$ is the integrated extinction in the V-band along the line
of sight $(l_i,b_i,r_i)$ in galactic coordinates.\\
The extinction decreases strongly with the galactic latitude.
For this reason, an opacity exponential model has been chosen :
$$
\rho_{op}(l,b,r)=A_0(l,b,r)\exp\left(-\frac{|r\sin(b)|}{h_0}\right)
$$

$A_0(l,b,r)$ represents the extinction fluctuations around an exponential
mean extinction law at the point ($l$,$b$,$r$)
and $h_0$, the characteristic height of the extinction structures. The
value of $h_0$ is a parameter which is determined during the inverse
process. The computed value of $h_0$ is 120 pc that is consistent
with previous studies (e.g. Sharov, 1964).
Since the sample of clusters is limited to a finite number of
lines of sight that are relatively spread-out, we have assumed
some regular properties of $\rho_{op}(l,b,r)$ and introduced 
a smoothing length which
corresponds to the mean inter-clusters distance. The details
of the absorption clouds, with a length lower than the correlation length,
are not detected. A 300 pc correlation length gives us global
fluctuation extinction tendencies. 
In Fig. 5 and Fig. 6, 
we show the extinction structure in the galactic plane 
and in the cross section of the galactic plane, respectively. 
These figures show that the absorption in the galactic plane is 
highly variable from almost zero to 1.5 mag/kpc, no information
are available at high galactic latitude.

Fig. 7 shows the histogram of $\delta$ value ($\delta =
\frac{Av(obs)-Av(model)}
{\sigma_{Av(obs)}}$), and in Figure 8 we plot the observed extinction Av(obs) 
against that derived from our model Av(model).
From Figure 8, we did not see an overestimation A$_{v}$(model) for small
distance ($r$ $<$ 500 pc), which seems to be the most significant 
improvement in our model over AGG92's model (see Figure 4).
The  error distribution in Figure 7 is asymmetrical and the negative 
tail shows that some extinction values are overestimated. This could be 
explained by the  
fact that the model does not take into
account the fluctuations at the smallest scales and does not represent
windows without extinction.  From Figure 7 and Figure 8, we can also see that 
model underestimates some extinction values at large Av(obs). 
Recently, Hakkila et al. (1997) found that large-scale extinction studies 
tend to underestimate extinction at distances $r$ $>$ 1 kpc. This is in 
agreement with our result.  For this reason, in the following sections, 
we limit our studies for $r$ $<$ 1000 pc.

In order to establish an extinction model in the solar neighborhood,
we have divided the  galactic plane into 36 cells with
$\Delta l=10^o$. At the center of each cell, the extinction curve 
has been computed. Figure 9 shows the $A_{v}$ as a function of $r$ for each
of the 36 cells.
We have compared Figure 9 with that of Neckel \& Klare (1980) and 
Pandey \& Mahra (1987), and   
found that, although A$_{v}$(r) changes from one cell to another
significantly, 
the scatter in the A$_{v}$(r) is in agreement with 
that 
in the diagrams of Neckel \& Klare (1980) and Pandey \& Mahra (1987). 
By comparing our results with those of AGG92, 
we can see that the general trends are similar:
strong
 extinction values in the longitude interval $20^o-40^o$ and
week extinction values in the interval $210^o-240^o$. However, 
we have observed  the presence of a strong extinction in the $350^o-360^o$ 
interval
at about 1.5 kpc which is not  evident in AGG92's 
model.

For each cell, an analytic polynomial expression is used to fit the results 
from the inverse method within r $<$ 1000 pc :
\begin{equation}
E(l)=a_1 r + a_2 r^2 + a_3 r^3 + a_4 r^4 
\end{equation}
where $E(l)$ is the extrapolated extinction at the longitude $l$ which
belongs to  the interval $[i\Delta l, (i+1)\Delta l]$ and  $r$ is the
distance
in kpc.  The coefficients $a_1,a_2,a_3$ and
$a_4$ are given in Table 1.

In order to provide a formal precision on our extinction model, 
we have made some comparisons with the observations for 155 stars from 
Neckel \& Klare (1980).  The stars selected have a spectral type B, A or F. 
O-type stars are rejected because they could have circumstellar extinction. 
In Figure 10, we plot the observed extinction A$_{v}$(obs) 
against that derived 
from our model predictions A$_{v}$(model). 
Considering the error on extinction the agreement is 
satisfactory.
We have grouped the Neckel \& Klare sample into four groups 
according to  A$_{v}$(model).  
In Figure 11, we show the histograms of Av(obs) in each group. 
Since the sample is small in each group, 
the errors ($\sigma_{Av}$)  have been derived by  
a robust estimator (Morrison \& Welsh 1989; Chen 1997a). 
Table 2  
shows the results.

\begin{table}
\caption[]{\em The extinction error ($\sigma_{Av}$) in each group Av(model)}
  \begin{center}
    \begin{tabular}{cccc}
      \hline
$Av(model)$  &     $\overline{Av(model)}$ &  $\sigma_{Av}$  &  number of
stars \\
\hline
0-0.25   &            0.15      &         0.20   &        60 \\
0.25-0.5        &  0.35      &         0.27     &     31 \\
0.5-1.0        &   0.75      &         0.56    &      42 \\
1.0-1.5        &   1.21       &         0.57     &      22 \\
    \end{tabular}
  \end{center}
\end{table}

Where $\sigma_{Av}$ is the total extinction error on A$_{v}$.
As a first approximation we neglect the contributions of the 
photometric errors in the Neckle \& Klare data, 
we  found $\sigma_{Av} (model)$  = 0.17 + 0.38 Av(model), which 
gives an approximated 
estimation of the external error in our extinction model.

\begin{figure}
      \mbox{}
      \vspace{14cm}
     \includegraphics{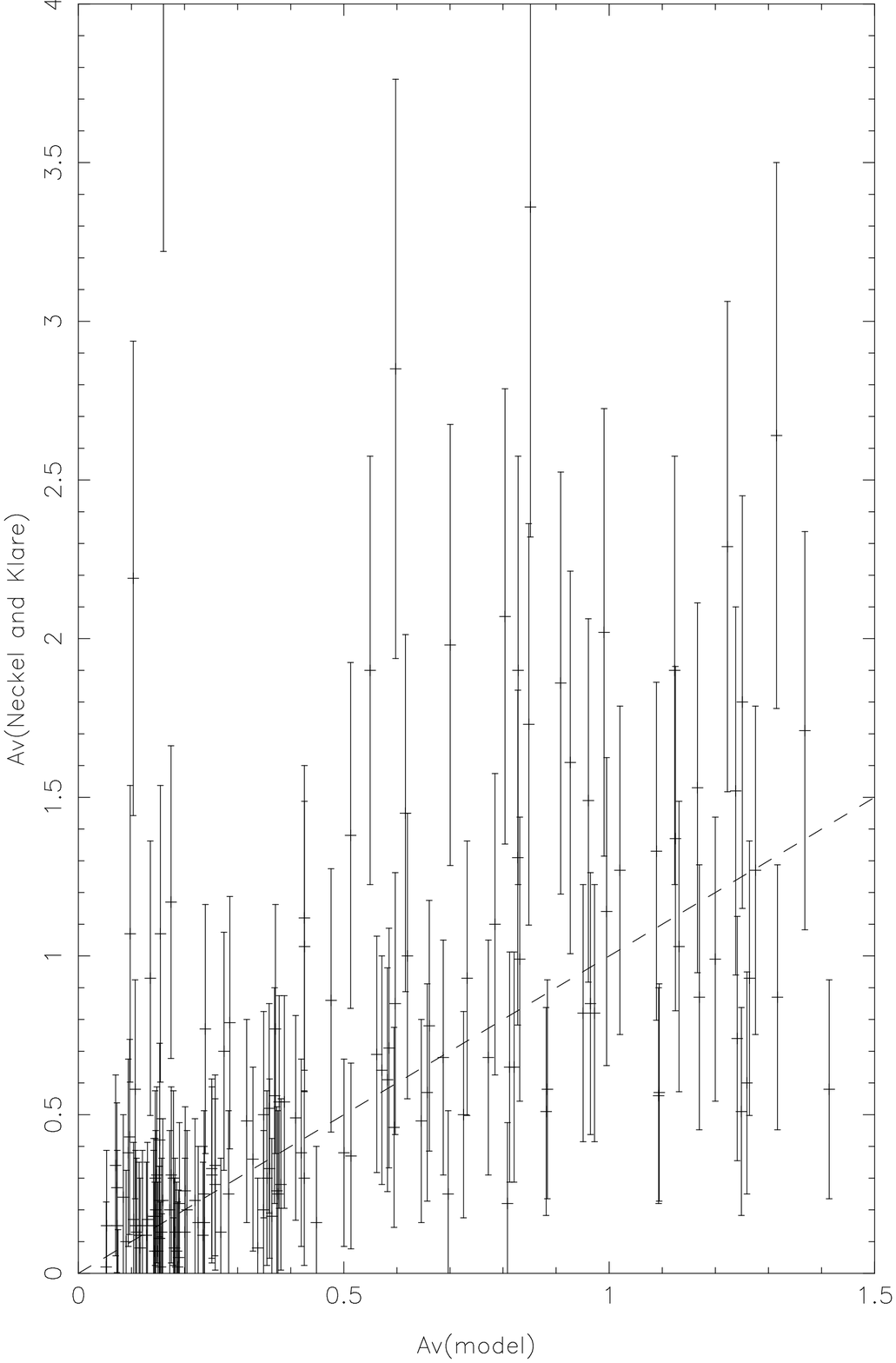}
  \caption[]{ The extinction derived from our model against the observations
for 155 stars from Neckel \& Klare (1980)}
   \end{figure}

For $\mid$b$\mid$ $>$ 10$^o$ we adopt 
the Sandage absorption law, which is widely used through the literature.
Finally, our extinction law in solar neighborhood is constructed as follows,
\\
\\
$A_v(r,l,b)$ = 0 ~~~~~~~~~~~~~~~~~~~~~~$\mid b \mid > 50$ 
\\
$A_v(r,l,b)$ = $0.165(1.192 - \mid tan b \mid )  \mid csc b \mid  
[1-exp(-r  \mid  sin b  \mid h_0^{-1})]$ ~~~~ $10 < \mid b \mid \le
50$
\\
$A_v(r,l,b)$ = E(l) ~~~~~~~~~~~~~~~$\mid b \mid \le 10$

Zero-reddening value in Sandage's model for $\mid$ b $\mid$ $>$ 50 has 
been criticized by de Vaucouleurs \& Buta (1983), and Lu et al. (1992), who found 
a polar extinction of 0.1 to 0.2 mag in A$_{B}$. 
From the color excesses derived from Str\"{o}mgren photometry for a complete 
sample of A3-G0 stars brighter than B = 11.5 and $\mid$ b $\mid$ $>$ 70, 
Knude (1996) found that 
the North Galactic Pole exhibits a complex reddening distribution. However, 
we found that Sandage's model with $\mid$b$\mid$ $>$ 50 is in agreement 
with the Burstein \& Heiles maps (1982). Moreover, Hakkila et al. (1997) 
compared several previous results in the directions of the Galactic poles
(see their Table 3), and found an average extinction A$_{v}$ = 0.1 $\pm$ 
0.2 mag in both galactic poles. We believe that Sandage's model in high  
galactic latitudes is a good approximation to observations.

\section{Comparison with Hipparcos observations}

\subsection{Hipparcos observations}
Hipparcos provides an immense quantity of accurate astrometric and 
photometric  data from which many branches of astrophysics 
are benefiting. 
The primary result is an astrometric catalogue of 118218 entries 
nearly evenly distributed over the sky 
with an astrometric precision in position, proper 
motion and parallax of 1 mas or mas/yr, or better for the brightest stars.
More details can be found in 
the comprehensive introduction of the catalogue (Perryman, 1997, ESA, 1997).
Hipparcos observations  include a survey, which is a complete
sample selected according to a combined magnitude-spectral class
-galactic latitude criterion. The selection  criterion
is V $\le$ 7.9 + 1.1 $\mid$ sin(b) $\mid$ for stars earlier
than or equal to  G5, and V $\le$ 7.3 + 1.1 $\mid$ sin(b) $\mid$
for later spectral type stars.

In this paper, we have used a complete sample with V $\le$ 7.3 mag.
Early-type stars are known to  be associated strongly with the Gould
belt and the moving groups (Chen et al. 1997). 
The non-uniform distribution of early type stars
is not included in our Galaxy model. Therefore we have excluded the O, B and
earlier A-type stars with B-V $\le$ 0.0. 
With this selection criterion we are left with 18072 stars
with V $\le$ 7.3 mag and B - V $>$ 0 from the Hipparcos catalogue.

\subsection{Galactic model}

A Galaxy structure and kinematic model has been constructed 
(Chen 1997b). The model 
includes a thin disk, a thick disk, and a halo, which can predict the 
magnitudes, positions, colours, proper motions, radial velocities and 
metallicities according to the selection criteria used in the observations.  
Since disk stars are dominating in Hipparcos survey, we simply 
describe the main characteristics of the disk population in our Galaxy model. 
Any further detail of the model can be found in Chen (1997b).

The stellar density laws used for the disk population are exponential.
For the distribution of stars 
perpendicular to the plane of the Galaxy, which is known to  vary with
luminosity, 
we adopted  the relations derived by Bahcall \& Soneira (1980).
The luminosity function for the disk stars given by Wielen et al. (1983) 
has been used in our model. 
The variation of the velocity dispersion in the solar neighborhood 
with spectral type and 
luminosity class has been adopted from Delhaye (1965) and Ratnatunga 
et al. (1989). 
We have adopted a velocity dispersion gradients from the 
Galactic plane suggested by Fuchs \& Wielen (1987).
Mendez \& van Altena (1996) have derived the distance of the Sun 
from the Galactic plane, Z = 2 $\pm$ 34 pc and Z = -8 $\pm$ 19 pc, from 
two fields. But, recently they (Mendez \& van Altena, 1998) found 
a distance of the Sun from the symmetry plane of the Galaxy of Z = 27 $\pm$ 3 pc.
Using a large sample of OB stars within 10 degrees of the galactic plane,
Reed (1997) has investigated the Sun's displacement from the galactic plane and 
found  Z = 10 $\sim$ 12 pc.  We found that this parameter is not well determined, 
for this moment, we keep Z = 0 in our model. 

In order to make comparisons with the Hipparcos observations, 
an area integration 
was performed by adding  up the contributions of many regions nearly 
uniformly distributed over the sky. A grid with spacings of 5$^o$ 
in latitude and 15$^o$ in longitude, divided the sky into 562 regions each 
of 75 deg$^{2}$.  Longitude spacings at high latitude are increased from the 
value selected at the Galactic equator to keep the areas in each cell 
approximately constant.

\begin{figure*}
      \mbox{}
      \vspace{15cm}
     \includegraphics{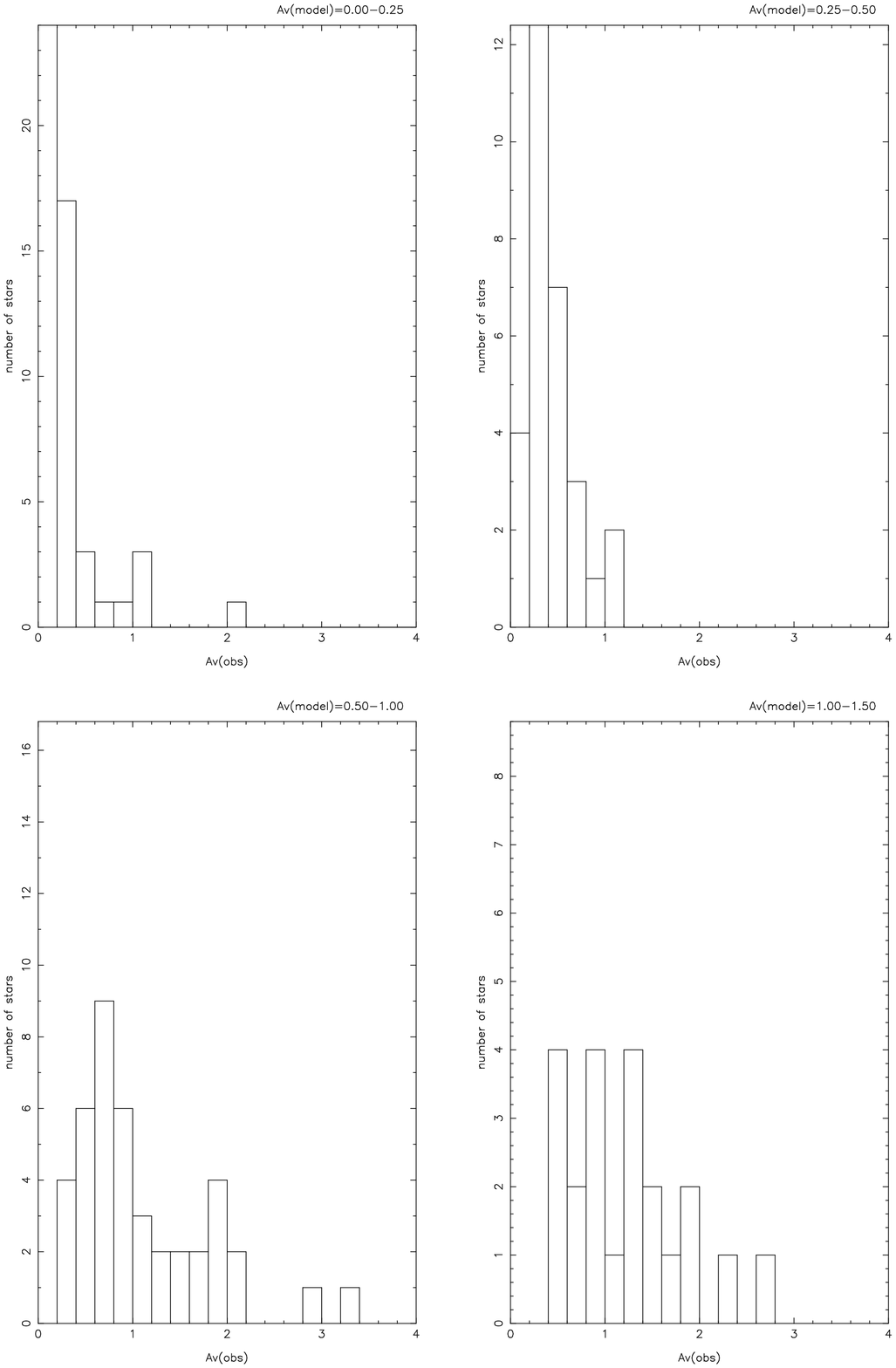}
  \caption[]{ The histogram of Av(obs)}
   \end{figure*}

\begin{figure*}

      \mbox{}
      \vspace{12cm}
     \includegraphics{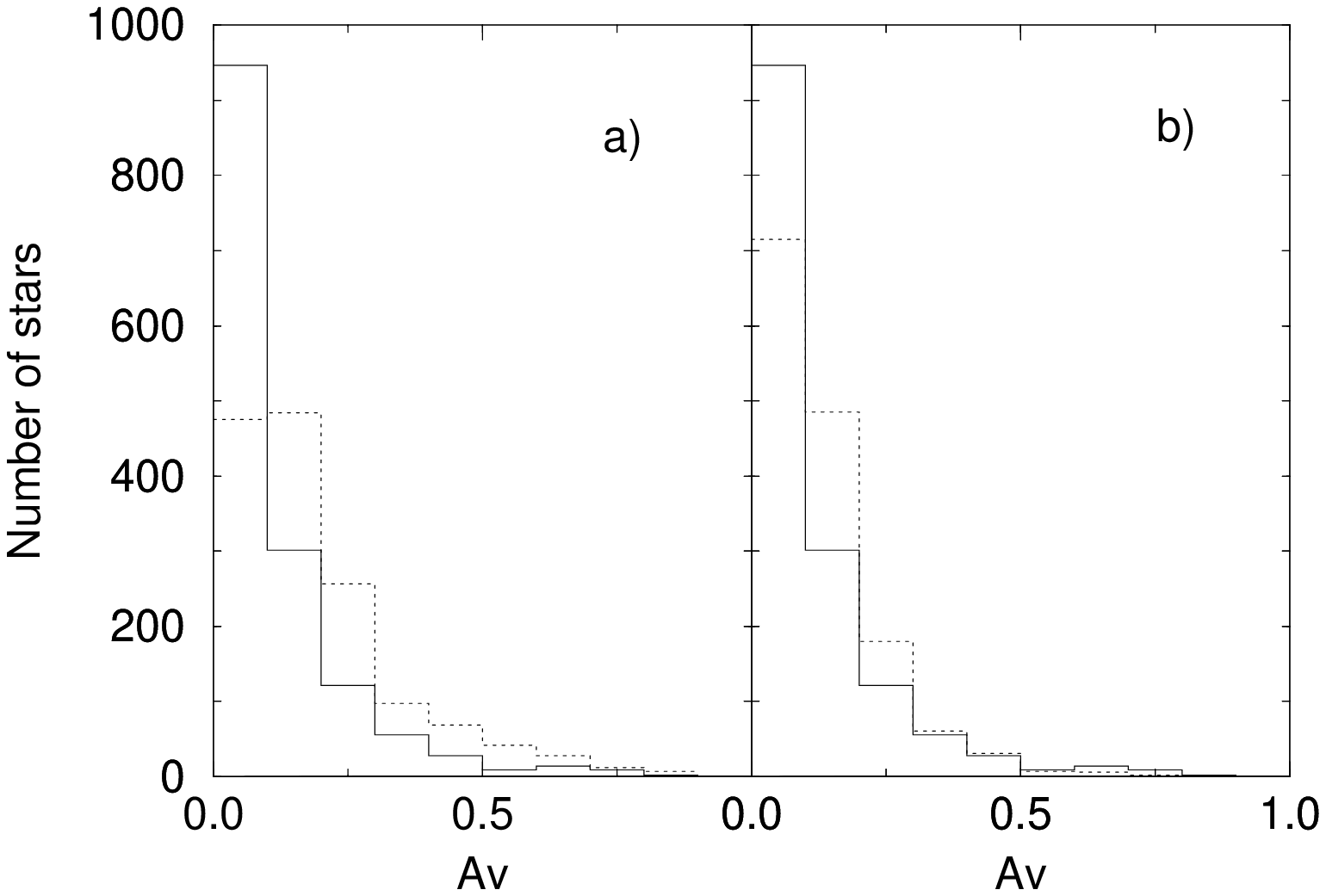}
  \caption[]{ Comparison between the observed A$_{v}$ distributions
and the model predictions, a): AGG92 model has been adopted, b):
Our model has been adopted}
   \end{figure*}

\begin{figure*}

      \mbox{}
      \vspace{17cm}
     \includegraphics{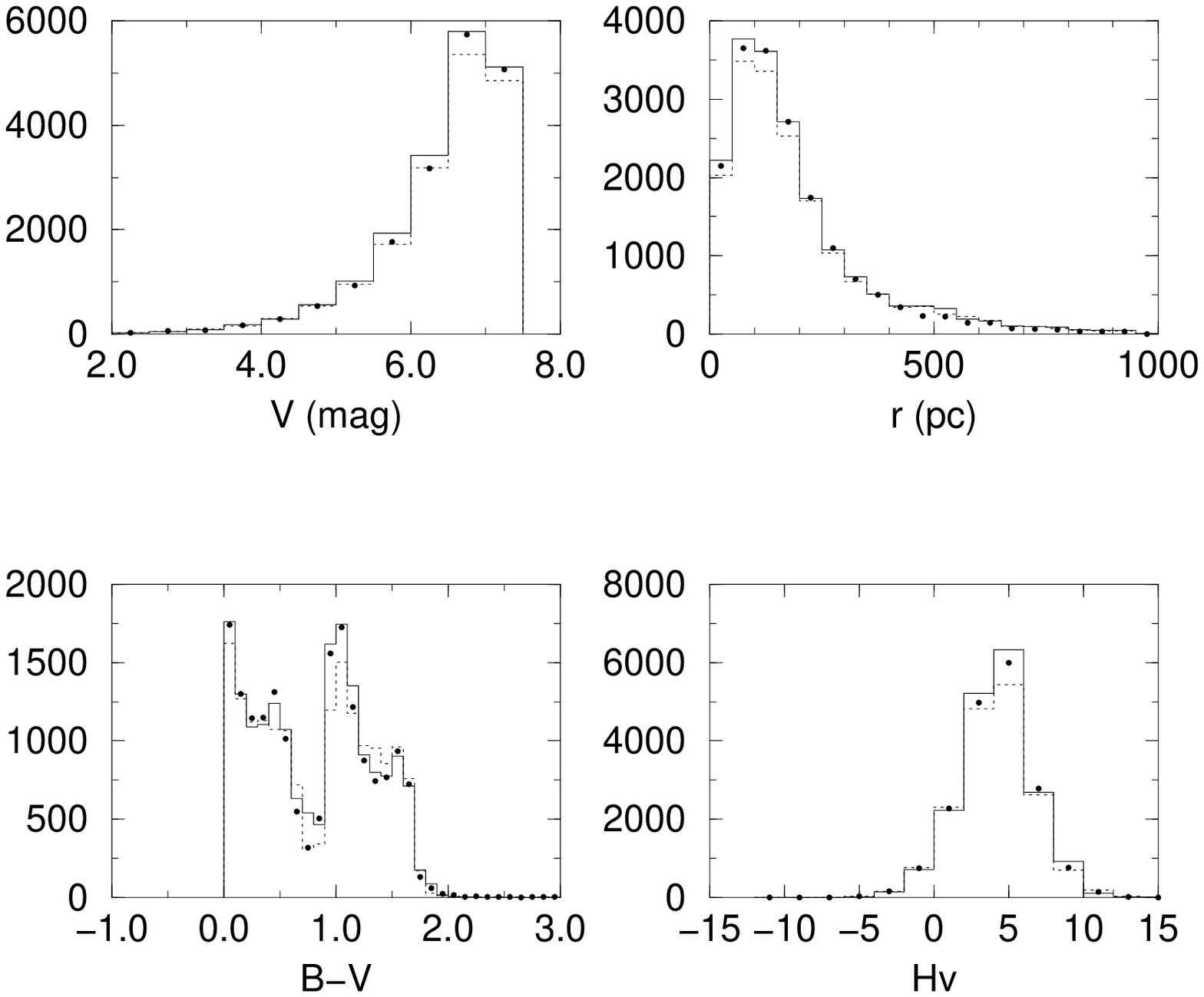}
  \caption[]{Predicted vs. Observed ({\sl circles}) distributions
  for whole sky. The solid line indicates our model predictions,
while the dotted line indicates the prediction from AGG92's model.}
   \end{figure*}

\subsection{ Comparisons of the model with a selected sample 
from Hipparcos data}
Domingo (1998) has compiled a sample of  B, A and F-type stars
from Hipparcos observations with uvbyH$_\beta$ photometry. 
Distances have been derived from Hipparcos parallaxes, and the 
Str\"{o}mgren photometric data come from the Hauck and Mermilliod 
(1990) compilation and the new observations performed by 
Barcelona group (Figueras et al., 1991, Jordi et al., 1996). 
In order to derive reliable physical parameters, we have not included 
in our sample any stars known to be or suspected of being variable,
spectroscopic binary, known as peculiar (Am, Ap, $\delta$ Del, ...) and 
those with variable radial velocity. Stars belonging to double or multiple 
systems for which only joint photometry is available have also been rejected. 

The absorption has been derived from Str\"{o}mgren photometry (Domingo, 1998). 
We have selected stars in our sample with A$_{v}$ $\ge$ 0 and $r$ $<$ 1000 pc.
Our sample includes 1492 stars and has an average distance {$\bar{r}$} = 
177 pc.
In Figure 12, we have compared AGG92 and our extinction model with 
the observations.
The average extinction derived from Str\"{o}mgren photometry in the sample 
is 0.121 $\pm$ 0.003 mag, and the average 
extinctions derived from AGG92 and our extinction model are 
0.203 $\pm$ 0.005 mag and 
0.127 $\pm$ 0.003 mag, respectively. 
It can be seen that AGG92 model provides a systematically larger 
extinction, and our model gives a better fit to observations.   
This is in good agreement with our result from open cluster dataset.

\subsection{ Comparisons of the model with Hipparcos data from whole sky} 

The predictions from our and AGG92 extinction models
have been compared with the Hipparcos observations in Figure 13.
The reduced proper motion $H_v$ is defined according to Luyten (1922): 
\begin{equation}
H_v = V + 5\log \mu + 5
\end{equation}
where $\mu$ is the total proper motion expressed in arcsec yr$^{-1}$.
With our extinction law, the model predicts 18486 stars, 2.3$\%$
more than that from the Hipparcos observations, while with the
Arenou et al. (1992)
extinction law, the model predicts 17229 stars, 4.7$\%$ less than
that from real observations.
The main features of the observed  distribution, for example,  the two
peaks near B-V = 0 and 1.0 mag, are present in the models and have
the correct amplitude.
In order to compare different extinction laws with observations in the 
4-dimensional space of ($V$, $B-V$, $H_v$, $r$), we use a cluster analysis 
algorithm (Chen 1996) to choose the best fit model. 
The Hipparcos data, made of magnitude ($V$), colour ($B-V$), 
reduced proper motion ($H_v$) and distance ($r$), is merged with the simulated 
sample. We carry out cluster analysis for this merged sample,     
then, we separate the observed data from the simulated one 
and compare the distribution between the model predicted stars and the real 
observed stars. The content of clusters can be interpreted from the model 
predictions. Table 3 gives the main physical parameters of each cluster
from our model predictions, and Figure 14 shows the positions 
of each cluster in H-R diagram.

\begin{table*}
  \caption[]{\em The means and standard deviations (between brackets) of 
observables and physical parameters in each cluster together with the 
cluster size. The physical parameters (M$_{v}$, 
$\sigma_{U}$:$\sigma_{V}$:$\sigma_{W}$) are derived by the corresponding 
simulated stars }
  \begin{center}
    \begin{tabular}{ccccccc}
      \hline
  & V &M$_{v}$    & $B-V$  & $r$ & $\sigma_{U}$:$\sigma_{V}$:$\sigma_{W}$ & N \\
  & mag & mag & mag & pc & km/s &stars \\
\hline
Cluster 1   & 7.0    & 1.3     & 0.3    & 154 & 21:14:13 & 3948 \\
            & (0.21)  & (1.64) & (0.23) & (96.2) & & \\
Cluster 2   & 4.0    &  0    & 1.0   & 80.4 & 34:21:17 & 782 \\
            & (0.89)  & (1.83) & (0.54) & (66.1) & & \\
Cluster 3   & 6.4   &   2.3   & 0.5   &84.7  &27:19:16  & 3049   \\
            & (0.33)  & (1.73) & (0.29) & (58.3) & & \\
Cluster 4   & 6.9    &  0.67     & 1.15   & 182 & 37:28:20 &3239    \\
            & (0.29)  & (1.27) & (0.34) &(100.8) & & \\
Cluster 5   & 5.3    &  1.7    &  0.4  & 62  &25:17:14   & 1535 \\
            & (0.47)  & (1.67) & (0.35) & (44.7) & & \\
Cluster 6   & 5.8    & -0.4    &  1.5  &  178 & 37:24:19  & 2357  \\
            & (0.44)  & (1.20) & (0.29) & (84.0) & & \\
Cluster 7   & 6.9    & -1.4    &  1.6  & 420 & 42:29:23  &  3162 \\
            & (0.35)  & (0.98) & (0.39) & (163.4) & & \\
\hline
    \end{tabular}
  \end{center}
\end{table*}

The $\chi^{2}$  statistics is used to test the capability of models to
represent the data and  choose the best fit to the observation:

\begin{equation}
\chi^{2} =\sum_{i=1}^{k} \frac {(N^{i}_{obs}-N^{i}_{model})^{2}}
{ N^{i}_{model}}
\end{equation}

where $N^{i}_{obs}$ and $N^{i}_{model}$ are the number of stars in cluster
$i$ in observed data and simulated data, respectively.  $k$  is the
total number of clusters, we found that $k$ = $7$ is good enough for 
this comparison (Chen 1996).

Obviously, if the model is a good representation of the observations, then
it should produce frequencies comparable with the observed ones
in each cluster.
A large value of $\chi^{2}$ indicates that the null hypothesis is rather
unlikely. From the sample, we have derived $\chi^{2}$ = 28 and 43 
for our extinction model and AGG92's model. 
The method (called as chi-by eye)  described above has been tested by 
Monte Carlo simulations and used to select the best-fitting 
model (Chen 1996). This 
method has also been used by Ojha et al(1996) to constrain galactic structure 
parameters.  
Our results show  that
our extinction model gives a better fit to Hipparcos observations. 
We should point out that in our and AGG92 extinction model, 
the probabilities that the model reflects
the reality are ruled out 
at 3.4 and 4.8 
sigma Gaussian equivalent level.  
This is due to the fact that the statistics of the errors in the data is not a 
Poisson statistics, because of some systematic errors in the 
photometry and astrometry in our Galaxy model, 
such as disk colour-magnitude diagram, 
absolute magnitude calibration for giants, 
velocity distribution and density law of the galactic disk stars, and the 
little fluctuation of the interstellar clouds as well. 
In this paper, for a given Galaxy model (Chen 1997b), 
we compare model predictions from different extinction laws 
with observations, which can  give us 
some information  about the uncertainty in the simulated sample 
due to the extinction.
In a forthcoming paper, we plan to improve galactic structure and kinematical 
parameters  
from Tycho catalogue with the help of our Galaxy model and extinction law. 

\begin{figure}

      \mbox{}
      \vspace{9cm}
     \includegraphics{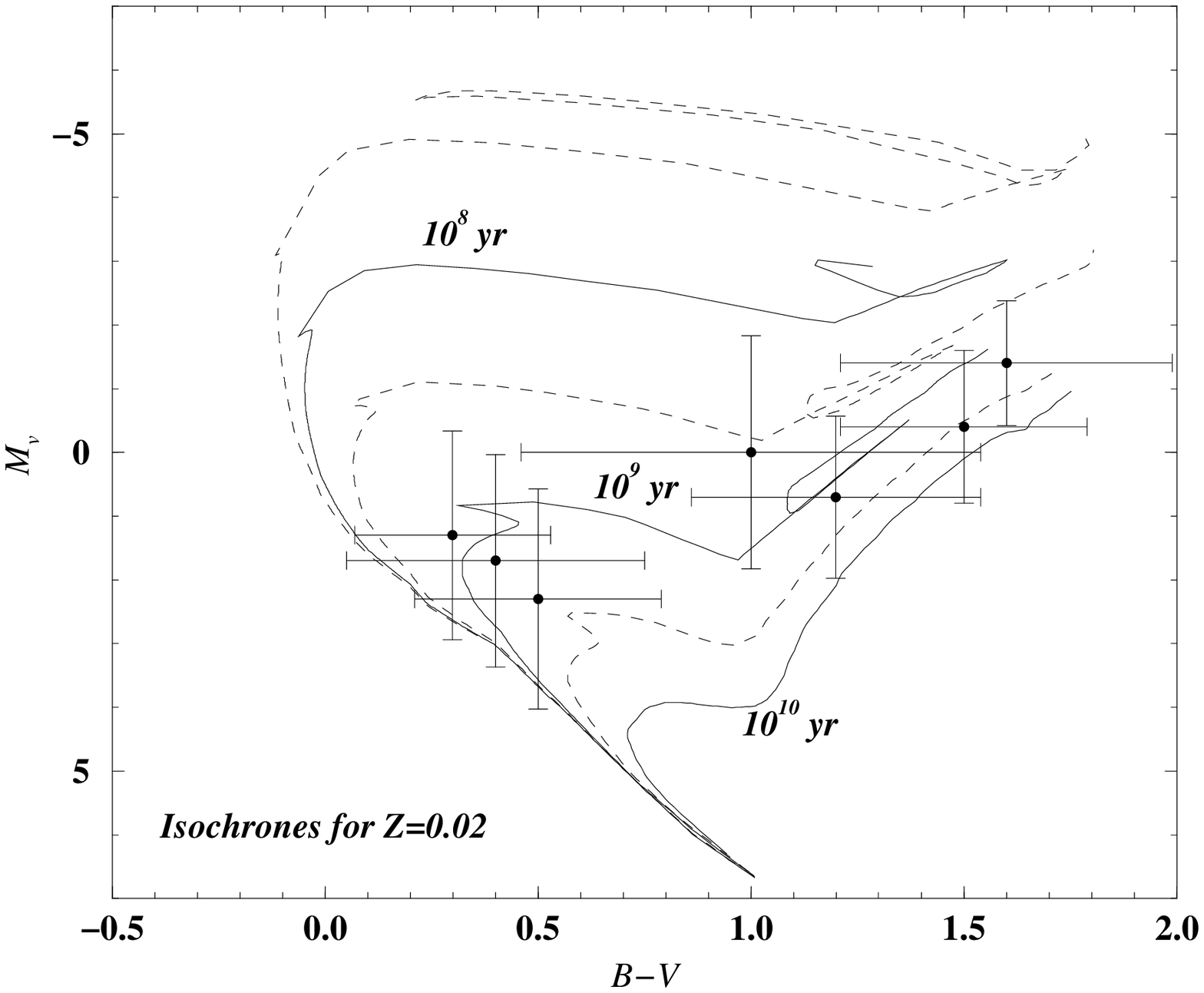}

  \caption[]{ The means and standard deviations of 7 clusters.
 Isochrones from Schaller et al. (1992) for solar metal content
  have been overplotted. Considering that our sample has an average
  extinction A$_{v}$ = 0.2 to 0.3 mag, B-V for Isochrones has been
  shifted 0.1 mag
  from (B-V)$_{o}$.}
   \end{figure}

 \begin{figure}

      \mbox{}
      \vspace{11cm}
     \includegraphics{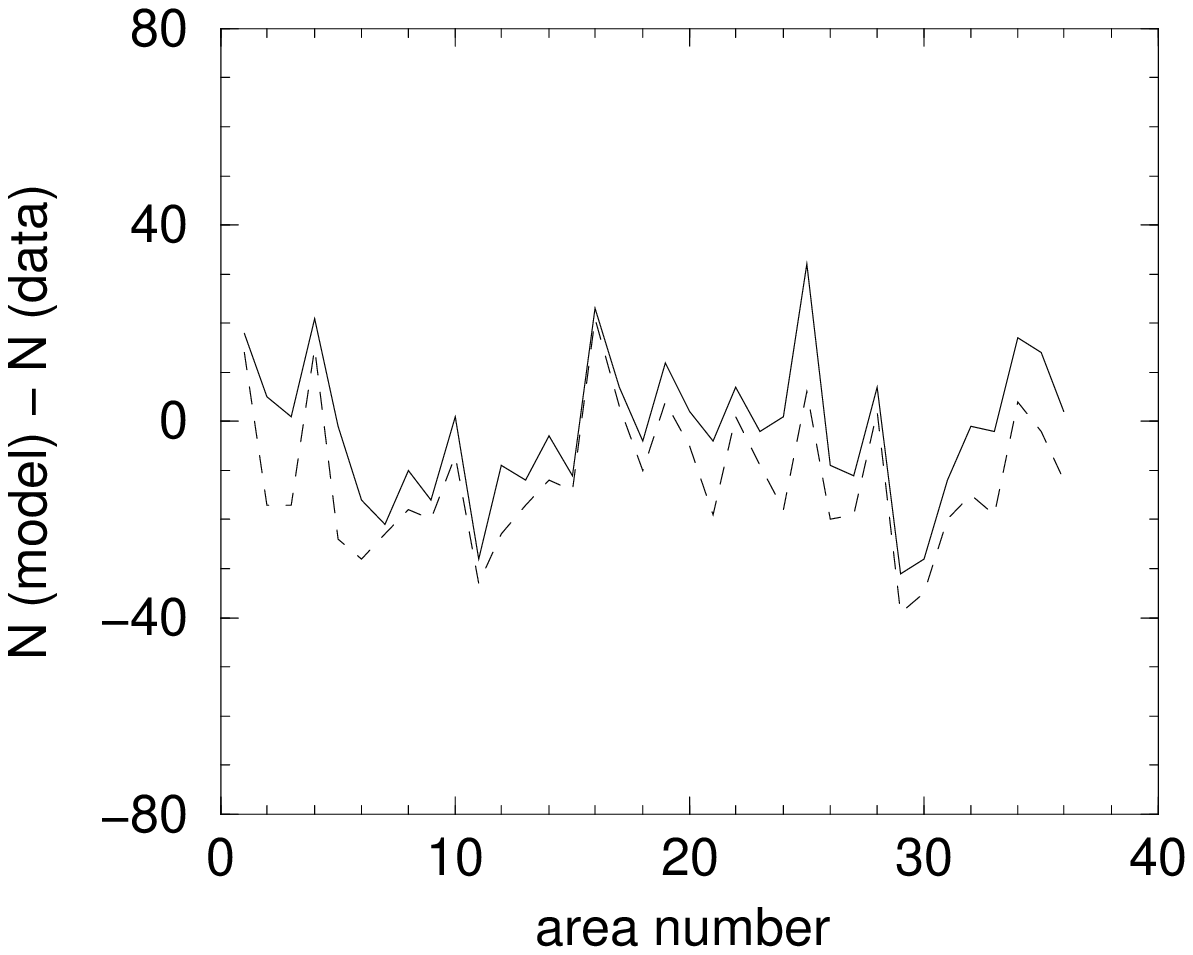}

  \caption[]{The difference $\delta = N (obs) - N (model)$ as a function
 of the cell number (from 1 to 36). The solid line indicates our of the cell number (from 1 to 36). The solid line indicates our model
   predictions, while the dotted line indicates the prediction
  from AGG92's model}
   \end{figure}

\subsection{ Comparisons of the model with Hipparcos data at    
low galactic latitudes}

In this section, we compare the model predictions with
Hipparcos observations at low galactic latitudes 
($\mid$ b $\mid$ $<$ 10$^{o}$).
Since the stellar density is about 5.5 stars per square degree in the 
galactic plane from Hipparcos observations,
comparisons must involve relatively large areas in the sky. 
We have used 36 areas,  each of
them covers  from -10$^{o}$ to 10$^{o}$ in galactic latitude and 10$^{o}$ in
galactic longitude.

\begin{table*}
  \caption[]{\em The number of stars in each cell derived from 
Hipparcos observation ($N_0$) and predicted by AGG92's extinction law 
($N_1$) and our extinction law ($N_2$).} 
  \begin{center}
    \begin{tabular}{cccccccc}
Range of $l$  & $N_{0}$ & $N_{1}$ & $N_{2}$  
&Range of $l$  & $N_{0}$ & $N_{1}$ & $N_{2}$\\
 (degrees)  & (stars) & (stars)  & (stars)
 & (degrees)  & (stars) & (stars)  & (stars)
 \\
\hline 
0-10  & 97 & 111 & 115    & 180-190  &103  &107  &115 \\ 
10-20  & 104 & 87 & 109   & 190-200  &119   &114 &121 \\
20-30  & 104 & 87 & 105   & 200-210  &135   &116   &131  \\ 
30-40  & 91 & 106 & 112    &210-220  &107  &108  &114 \\
40-50  & 121 & 97 & 120    &220-230  &114   &105  &112 \\
50-60  & 131 & 103 & 115   &230-240  &125  &107 &126 \\
60-70  & 142 & 119 & 121  & 240-250  &95 &101  &127 \\
70-80  & 133 & 115 & 123  & 250-260  &137  &117  &128 \\
80-90  &  136 & 116 & 120  & 260-270 &132  &113   &121  \\
90-100  &  116& 109 & 117  & 270-280  &112 &114 &119 \\
100-110  & 151 & 118 & 123 &  280-290 &152  &113  &121 \\
110-120  & 135 & 112 & 126 &  290-300 &150  &115  &122 \\
120-130  &  133& 116 & 121 &  300-310  &132  &112  &120 \\
130-140  & 120 & 108 & 117 &  310-320   &124  &109  &123 \\
140-150  &  127& 113 & 116 &   320-330  &126  &107  &124 \\
150-160  &  80& 101 & 103  &  330-340   &101  &105  &118 \\
160-170  &  100& 103 & 107 &  340-350   &109  &107  &123 \\
170-180  & 123 & 113 & 119 &  350-360   &117  &105  &119 \\
      \end{tabular}
  \end{center}
\end{table*}

AGG92's extinction law and 
the law presented in this 
study have been used 
for comparisons  with the observations. 
In Table 4, we show the number of stars in each cell derived from Hipparcos 
observation ($N_0$) and predicted by AGG92's extinction law ($N_1$) 
and our extinction law ($N_2$). 
We have used the $\chi^2$ statistics (eq. 6) to compare the model 
predictions and observations.  
With our extinction law, we found $\chi^2$ = 62.3 and for AGG92's model, 
$\chi^2$ = 108.7. The model predictions are at 2.6 and 5.7 sigmas from the data. 

In Figure 15, we plot the difference $\delta = N(obs)-N(model)$ 
as a function of the cell number (from 1 to 36). 
From $\chi^2$ and Figure 15, we can see that our extinction law provides 
a better 
fit to Hipparcos observations than  AGG92's one, which predicts less
stars than observed.
On the another hand, we found that both models predict too less stars 
between $l$ = 50 to 130. From Table 4, we can see there are 
152 stars between $l$ = 280 to 290, and only 80 stars 
between $l$ = 150 to 160. 
Neither our extinction law nor AGG92's model can model this large change.
A more detailed absorption law is needed for modeling the extinction 
substructure in the galactic plane.

\section{Discussions and Conclusions}

At low galactic latitudes ($\mid$ b $\mid$ $<$ 10$^{o}$), the absorptions
are patchy and not well known.
In this paper colour excesses and distances of open clusters compiled 
by Mermilliod (1992) have been used to study the interstellar extinction.
We have compared our extinction model with AGG92 and Hipparcos observations.
The main results of this paper can be summarized as follows.

1. An inverse method (Tarantola \& Valette, 1982)  was 
used to construct the extinction map in the galactic plane.
We found the lowest absorption  
between $l$ = 210$^{o}$ to 240$^{o}$, and the 
highest  
absorption between $l$ = 20$^{o}$ to 40$^{o}$.

2. By using three different samples,  we 
found that AGG92's results are higher than the observations 
for small distances. 

3. We have constructed an extinction law by combining Sandage extinction law 
with $\mid$ b $\mid$ $>$ 10$^{o}$ and our extinction law in the galactic plane. 

4. AGG92 and our extinction models have been compared with a well selected 
sample from Hipparcos observations with Str\"{o}mgren photometry. We found that 
AGG92's model provides a slight overestimation of A$_{v}$.

5.  Galaxy model predictions with our and AGG92's  
extinction laws have been compared with the Hipparcos observations for 
the whole sky and at low galactic latitudes, respectively. 
The main characteristics in the observed distributions 
(for example, two peaks in $B-V$) are well predicted by both models. 
The $\chi^2$ statistics, used to choose the best fit to Hipparcos 
observations, shows that our extinction model provides a better fit to 
data.

6. Interstellar absorption in the galactic plane is highly variable 
from one direction to another. 
We believe that a more detailed 
absorption law, which can represent the extinction at smaller scales,
 is needed for modeling the
extinction  substructure in the galactic plane.

These results are very useful for us to  investigate the stellar kinematics 
in the solar neighborhood from Hipparcos and Tycho data in the near future.  
But a new and more detailed extinction model from Hipparcos observations 
is still missing.

\begin{acknowledgements}

We thank Drs. C. Jordi, J. Torra and S. Ortolani for helpful comments. 
Thanks to A. Domingo to provide us his sample before publication.
The comments of the referee, F. Arenou, which helped to improve the 
content of this paper, are gratefully acknowledged.
This research has made use of Astrophysics Data System Abstract Service 
and SIMBAD Database at CDS, Strasbourg, France.
This work has been supported by CICYT under contract PB95-0185 and 
by the Acciones Integradas Hipano-Francesas (HF94/76B).  
\end{acknowledgements}

{}


\begin{thebibliography}{}
\bibitem{} Arenou F., Grenon M., G\'omez A., 1992, A\&A 258, 104
\bibitem{} Bahcall J.N., Soneira R.M., 1980, ApJS 47, 357
\bibitem{} Berdnikov L. N., Pavlovskaya E.D., 1991, 
Sov. Astron. Lett., 17, 215
\bibitem{} Burstein D., Heiles C., 1982, AJ, 1165
\bibitem{} Carraro G., Ortolani S., 1994, A\&AS 106, 573 
\bibitem{} Chen B., 1996, A\&A, 306, 733
\bibitem{} Chen B., 1997a, AJ, 113, 311
\bibitem{} Chen B., 1997b, ApJ 491, 181
\bibitem{} Chen B., Asiain R., Figueras F., Torra J., 1997, A\&A 318, 29
\bibitem{} Chen B., Carraro G., Torra J., Jordi C., 1998, A\&A 331, 916 

\bibitem{} Crawford D.L., 1979, AJ, 84, 1858
\bibitem{} Delhaye J., 1965, in Galactic Structure, ed. A. Blaauw \& M. Schmidt
(Chicago: Univ. Chicago Press), 61
\bibitem{} de Vaucouleurs G., Buta R. 1983, AJ 88, 939
\bibitem{} Domingo A., 1998, Degree in Physics, Universitat de Barcelona
\bibitem{} ESA, 1997, The Hipparcos Catalogue, ESA SP-402
\bibitem{} Figueras F., Torra  J., Jordi C.,  1991, A\&AS  87,   319

\bibitem{} FitzGerald, M. P. 1968, AJ, 73, 983
\bibitem{} Friel E. D., 1995, ARAA 33, 381
\bibitem{} Fuchs B., Wielen R., 1987, in The Galaxy, ed. G. Gilmore \& 
B. Carswell (NATO ASI Ser. C, 207) (Dordrecht:Reidel), 375
\bibitem{} G\'omez A., Morin D., Arenou F., 1989, ESA-SP 1111, 23
\bibitem{} Hakkila J., Myers J. M., Stidham B. J., Hartmann D. H., 1997,
AJ, 114, 2043
\bibitem{} Hauck B., Mermilliod J.C., 1990, A\&A 96, 107

%\bibitem{} G\'omez A., Delhaye J., Grenier S., Jaschek C., Arenou F., 
%Jaschek M., 1990, A\&A, 236, 95
\bibitem{} Janes K., Adler D., 1982, ApJS 49, 425
\bibitem{} Jordi C., Figueras F., Torra J., Asiain R., 1996, A\&AS 115, 401
\bibitem{} Kaluzny J., 1994, A\&AS 108, 151
\bibitem{} Knude J. 1996, A\&A 306, 108
\bibitem{} van Leeuwen F., Hansen-Ruiz C.S., 
1997, in Hipparcos Venice'97 symposium, ESA SP-402, 643
\bibitem{} Lu N.Y., Houck J.R., Salpeter E.E. 1992, AJ 104, 1505
%\bibitem{} Lyng\a{a} G., 1987, "Catalogue of Open Cluster Data", $5$th edition,
%distributed by Centre de Donn\'ees Stellaires, Strasbourg, France
\bibitem{} Luyten W. J., 1922, On the relation between Parallax, Proper Motion,
and Apparent Magnitude (lick Obs. Bull. 336) (Berkeley: Univ. California press)
\bibitem{} Mendez R.A., \& van Altena  W. F., 1996, AJ 112, 655
\bibitem{} Mendez R.A., \& van Altena  W. F., 1997, A\&A in press
\bibitem{} Mermilliod J.-C., 1992, Bull. Inform. CDS n. 40, 115 
(June 1995 version)
\bibitem{} Mermilliod J.-C., Turon C., Robichon N., Arenou F., Lebreton Y.,
1997, in Hipparcos Venice'97 symposium, ESA SP-402, 643
\bibitem{} Morrison H.L., Welsh A. H., 1989, in Error, Bias and 
Uncertainties in Astronomy, edited by F. Murtagh and C. Jaschek 
(Cambridge University Press, Cambridge) 
\bibitem{} Neckel Th., Klare G., 1980, A\&AS 42, 251
\bibitem{} Ojha D., Bienayme O., Robin A. C., Cr\'ez\'e M., Mohan V., 
 1996, A\&A 311, 456 
\bibitem{} Pandey A.K., Mahra H.S., 1987, MNRAS 226, 635
\bibitem{} Perryman M.A.C., 1997, A\&A 323, 49
\bibitem{} Ratnatunga K.U., Bahcall J.N., Casertano S., 1989, ApJ 339, 106
\bibitem{} Reed, B. C., 1997, PASP, 109, 1145 
\bibitem{} Robichon N., Arenou F., Turon C., Mermilliod J.C., 
Lebreton Y., 1997, in Hipparcos Venice'97 symposium, ESA SP-402, 567 
\bibitem{} Sandage A., 1972, ApJ, 178, 1
\bibitem{} Schaller G., Schaerer D., Meynet G., Maeder A., 1992, A\&A, 96, 269 
\bibitem{} Sharov A. S., 1964, SvA, 7, 689 
\bibitem{} Tarantola A., Valette B., 1982, Rev. of Geo. and Space Physics,
20, 219 
\bibitem{} Turon C. et al., 1991, Database \& On-line Data in Astronomy,
eds. M.A. Albrecht and D. Egret, p. 67
\bibitem{} Valette B., Vergely J.L., 1998, in preparation
\bibitem{} Vergely J.L., Egret D., Ferrero R., Valette B., Koppen J., 1997,
in Hipparcos Venice'97 symposium, ESA SP-402, 603 
\bibitem{} Wielen R., Jahreiss H., Kruger R., 1983, in IAU Colloq. 76, the 
Nearby Stars and the Stellar Luminosity Function, ed. A.G. Davis Philip 
\& A. R. Upgren (Schenectady: L. Davis Press), 163 
\end{thebibliography}
\end{document}